\documentclass[epj,nopacs]{svjour}
\usepackage{graphics}
\begin{document}
\title{k-essence classical  Hamiltonian approach for an accelerated expansion of the Universe with $\omega\approx-1$}

\author{Somnath Mukherjee\inst{1}  % etc
 % - remove next line if not needed
%
}                     % Do not remove
\offprints{Somnath Mukherjee}
\mail{Dharsa Mihirlal Khan Institution[H.S],
            P.O:-New G.I.P Colony,  Dist:-Howrah-711112, India.}          % Insert a name or remove this line
\institute{Department of Physics, Dharsa Mihirlal Khan Institution[H.S],
            P.O:-New G.I.P Colony,  Dist:-Howrah-711112, India, \email{sompresi@gmail.com} }

\date{Received: date / Revised version: date}
\abstract{
We obtain lagrangian for  $k$-essence scalar field 
$\phi(r,t)$ with scalar curvature $k$ of Friedmann-Lemaitre-Robertson-Walker (FLRW) metric  . Obtained lagrangian has two generalised co-ordinates $\phi$ and logarithm of scale factor ($q=\ln a$). Classical Hamiltonian $({\mathcal H})$ is obtained in terms of two corresponding conjugate momentum $p_{q}$ and $p_{\phi}$. Solving Hamilton's equation of motion , we obtain classical solution for scale factor $a(t)$, energy density $\rho$, equation of state parameter $\omega$ and deceleration parameter $q_{0}$ .  At late time as $t\rightarrow\infty$, we have an exponential growth of scale factor with time, energy density $\rho$ becomes constant, which we can identify as dark energy density, equation of state parameter becomes $\omega\approx -1$ and deceleration parameter becomes $ q_{0}\approx -1$. All this results indicates an accelerated expansion of universe driven by negative pressure known as dark energy. 
\PACS{
      {PACS-key}{discribing text of that key}   \and
      {PACS-key}{discribing text of that key}
     } % end of PACS codes
} %end of abstract
\maketitle
\section{Introduction}
\label{intro}
From the observation of Type 1a Supernovae (SNe 1a) by The Supernova Cosmology Project \cite{Ref1,Ref2,Ref3,Ref4,Ref5,Ref6} and the High-Z-Supernova search team \cite{Ref7,Ref8,Ref9} it was first established that the universe is undergoing accelerated expansion. Recent observations with WMAP satellite \cite{Ref10,Ref11} and Planck satellite \cite{Ref12} also ensures an accelerated expansion of the universe, driven by negative energy now known as dark energy which accounts for 70 percent constituents of the universe.\\
Several cosmological model has been developed to understand the role of dark energy in the universe, out of which we have chosen $k$-essence model of scalar field $\phi(r,t)$ with non-canonical kinetic term $X$ as our field of study.\\  
The Lagrangian for the $k-$essence field is taken as \cite{Ref15,Ref16,Ref17,Ref18,Ref19,Ref20,Ref21,Ref22,Ref23,Ref24,Ref25,Ref26,Ref27,Ref28}
\begin{equation}
{\mathcal L}= -V(\phi)F(X)
\end{equation}
where
\begin{equation}
 X={{1\over2}{\partial_{\mu}{\phi}{\partial^{\nu}{\phi}}}}=\frac{1}{2}{\dot{\phi}^{2}}-\frac{1}{2}(\nabla\phi)^{2}
\end{equation}
Energy density $\rho$ for k-essence field is given by
\begin{equation}
\rho=V(\phi)[F(X)-2X F_{X}]
\end{equation}
with 
$F_{X}=\frac{\partial F}{\partial X}$.\\
and the pressure $P$ is given by
\begin{equation}
P={\mathcal L}=-V(\phi)F(X)
\end{equation}
 The conservation equation is given by
\begin{equation}
\dot{\rho}+3H(\rho+P)=0
\end{equation}
where $H$ is a Hubble parameter, defined in terms of scale factor $a$ as $H=\frac{\dot{a}}{a}$.
Considering homogeneity and isotropy of the universe $\phi(r,t)\approx \phi(t)$, so that $X=\frac{1}{2}{\dot{\phi}}^{2}$, we get from $(3)$,$(4)$ and $(5)$
\begin{equation}
(F_{X}+2XF_{XX})\ddot{\phi}+3HF_{X}\dot{\phi}+(2XF_{X}-F)\frac{V_{\phi}}{V}=0
\end{equation}

Considering constant potential ($V(\phi)$=constant, so that $\frac{\partial V}{\partial \phi}=0$) the conservation equation becomes
\begin{equation}
(F_{X}+2XF_{XX})\dot{X}+6HXF_{X}=0
\end{equation}
Solving this gives the scaling relation
\begin{equation}
XF_{X}^{2}=Ca^{-6}
\end{equation}
This is the scaling relation \cite{Ref22,Ref23} of $k$-essence cosmology. It plays a very important role in determining the lagrangian for the development of cosmological scenario of observational importance.
\section{ Lagrangian with curvature constant $k$}
Considering Friedmann-Lemaitre-Robertson-Walker (FLRW) metric of the form
\begin{equation}
 ds^{2}=-dt^{2}+a^{2}(t)[\frac{ dr^{2}}{1- kr^{2}}+r^2d{\theta}^{2}+r^2\sin^{2}\theta d{\phi}^{2}]
\end{equation}
where $a(t)$ is the cosmological scale factor and $k$ is the curvature constant that describes closed, flat and open universe for value $k=+1,0$ and $-1$.\\
 
From  Friedmann equation, we get :-
\begin{equation}
\frac{k}{a^{2}}+H^{2}=\frac{8\pi G}{3}\rho
\end{equation}
where $H=\frac{\dot{a}}{a}$ is the Hubble's constant and $\rho$ is the dark energy density.
Since we are considering constant potential, hence we can assume $V(\phi)=V_{0}$ as some constant. From $(3)$ and $(10)$ we get 
\begin{equation}
XF_{X}=\frac{1}{2}[F(X)-\frac{3}{8\pi GV_{0}}H^{2}+\frac{3k}{8\pi GV_{0}a^{2}}]
\end{equation}
Now considering equation $(8)$ and $(11)$ we get
\begin{equation}
F(X)=-2\sqrt{C}\sqrt{X}a^{-3}-\frac{3}{8\pi GV_{0}}H^{2}+\frac{3k}{8\pi GV_{0}}a^{-2}
\end{equation}
thus the k-essence lagrangian $(1)$ becomes
\begin{equation}
{\mathcal L}=-2\sqrt{C}\sqrt{X}V_{0}a^{-3}-\frac{3}{8\pi G}H^{2}+\frac{3k}{8\pi G}a^{-2}
\end{equation}
From the consideration of Homogeneity and isotropy of observed universe we will consider scalar field as $\phi(r,t)\approx\phi(t)$.\\
Now considering $q(t)=\ln a(t)$, $c_{1}=\frac{3}{8\pi G}$ and $c_{2}=\sqrt{2C}$ equation $(13)$ becomes
\begin{equation}
{\mathcal L}=-c_{1}{\dot{q}}^{2}-c_{1}ke^{-2q}-c_{2}V_{0}\dot{\phi}e^{-3q}
\end{equation}
This is the $k$-essence lagrangian with curvature constant $k$ of FRW metric and has two genaralised co-ordinate $q(t)$ and  $\phi(t)$. For flat universe with $k=0$, this reduces to the lagrangian Ref. \cite{Ref19}. In the  subsequent section we will use this lagrangian to obtain cosmological parameter of observational importance.
\section{Hamiltonian}
\label{sec:4}
In this section we will derive classical Hamiltonian corresponding to classical lagrangian $(14)$.\\ Conjugate momentum corresponding to $q$ is
\begin{equation}
p_{q}=\frac{\partial{\mathcal L}}{\partial \dot{q}}=-2c_{1}\dot{q}
\end{equation}
and conjugate momentum corresponding to $\phi$ is
\begin{equation}
p_{\phi}=\frac{\partial{\mathcal L}}{\partial \dot{\phi}}=-c_{2}V_{0}e^{-3q}
\end{equation}
thus the Hamiltonian corresponding to the $k$-essence Lagrangian 
\begin{equation}
{\mathcal H}=p_{q}\dot{q}+p_{\phi}\dot{\phi}-{\mathcal L}
\end{equation}
is obtained as
\begin{equation}
{\mathcal H}=-\frac{{p_{q}}^{2}}{4c_{1}}-\frac{c_{1}k{p_{\phi}}^{\frac{2}{3}}}{(c_{2}V_{0})^{\frac{2}{3}}}
\end{equation}
This is $k$-essence Hamiltonian. This is different from the earlier framed hamiltonian in the context that this has non zero curvature constant. Our objective is to develop cosmological parameters out of this Hamiltonian which will be consistent with the observed data.
\section{Scale factor a}
\label{sec:5}
Since $\phi$ is a cyclic co-ordinate thus $p_{\phi}$ is conserved.
Considering second term of Hamiltonian $(18)$ in terms of $q$
\begin{equation}
{\mathcal H}=-\frac{{p_{q}}^{2}}{4c_{1}}+c_{1}ke^{-2q}
\end{equation}
 Hamiltonian equation of motion for the generalised co-ordinate $q$ is determined as follows:-
\begin{equation}
\dot{q}=\frac{\partial{\mathcal H}}{\partial p_{q}}=-\frac{p_{q}}{2c_{1}}
\end{equation} 
\begin{equation}
\dot{p_{q}}=-\frac{\partial{\mathcal H}}{\partial q}=2c_{1}ke^{-2q}
\end{equation}
Considering equation $(20)$ and $(21)$  yields
\begin{equation}
\ddot{q}=-ke^{-2q}
\end{equation}
Since $q=\ln a$, this becomes
\begin{equation}
a\frac{d^{2}a}{dt^{2}}-(\frac{da}{dt})^{2}+k=0
\end{equation}
The solution  yields
\begin{equation}
a=Be^{c_{3} t}-kDe^{-c_{3} t}
\end{equation}
where $B$,$D$ and $c_{3}$ are constant. This is the scale factor with curvature constant $k$. For late time cosmology, as $t\rightarrow\infty$ , this becomes
 
\begin{equation}
a \approx Be^{c_{3} t}
\end{equation}
This shows the exponential increase of scale factor over cosmic time. 
\section{Energy density $\rho$}
Considering $(10)$ and $(24)$ we get
\begin{equation}
\rho=c_{1}kB^{-2}e^{-2c_{3}t}z +c_{1}{c_{3}}^{2}(1+kDB^{-1}e^{-2c_{3}t})^{2}z
\end{equation}
where $z=(1-kDB^{-1}e^{-2c_{3}t})^{2}$.\\
As we are considering late time cosmology, hence as  $t\rightarrow\infty$ energy density becomes
\begin{equation}
\rho\approx c_{1}{c_{3}}^{2}\approx constant
\end{equation}
This constant energy density can be identified as dark energy density. This is independent of curvature constant $k$.  Thus we see that energy density remains constant at late time independent of the nature of universe. We get this result out of $k$-essence classical Hamiltonian.
\section{Equation of state parameter $\omega$}
From Friedmann equation for the solution of the spatial part of Einstein's equation , the relationship betwen pressure $(P)$ and energy density $(\rho)$ is given by:-
\begin{equation}
-8\pi G P=2\frac{\ddot{a}}{a}+\frac{8\pi G}{3}\rho
\end{equation}
using $(24)$ this yields
\begin{equation}
P=-\frac{{c_{3}}^{2}}{4\pi G}-\frac{\rho}{3}=-\frac{2c_{1}{c_{3}}^{2}}{3}-\frac{\rho}{3}
\end{equation}
Thus equation of state parameter becomes
\begin{equation}
\omega=\frac{P}{\rho}=-\frac{2c_{1}{c_{3}}^{2}}{3\rho}-\frac{1}{3}
\end{equation}
since we are considering only late time cosmology, thus considering the form of $\rho$ at $t\rightarrow\infty$, we get
\begin{equation} 
\omega\approx-1
\end{equation}

This is consistent with the observational values \cite{Ref1,Ref2,Ref3,Ref4,Ref5,Ref6,Ref7,Ref8,Ref9,Ref10,Ref11,Ref12,Ref13,Ref14}. Thus we achieve negative equation of state through classical Hamiltonian approach. This also indicates the relevance of dark energy irrespective of nature of universe (closed, flat or open for $k=+1,0,-1$).

\section{Deceleration parameter $q_{0}$}
Deceleration parameter in terms of scale factor $a(t)$ is given by 
\begin{equation}
q_{0}=-\frac{\ddot{a}a}{{\dot{a}}^{2}}
\end{equation}
Considering the form of scale factor  $(24)$ this yields
\begin{equation}
q_{0}=-\left(\frac{1-kDB^{-1}e^{-2c_{3} t}}{1+kDB^{-1}e^{-2c_{3}t}}\right)^{2}
\end{equation} 
Considering late time cosmology, as $t\rightarrow\infty$, this becomes
\begin{equation}
q_{0}\approx-1
\end{equation}
This is consistent with the observational data, that shows the negativity of the deceleration parameter $q_{0}$ with values ranging from $-1\leq q_{0}\leq-0.5$ \cite{Ref3,Ref4}. This indicates an accelerated expansion of the universe, independent of type of universe $(k=+1,0,-1)$ closed, flat and open universe. This satisfies late time acceleration of the universe driven by negative pressure know as dark energy.\\
\section{Conclusion}
 Incorporating scaling relation $X{F_{X}}^{2}=Ca^{-6}$, $k$-essence lagrangian of non-canonical form ${\mathcal L}=-V(\phi)F(X)$ is developed into canonical form  with curvature constant $k$ of FLRW metric. It has two generalised co-ordinate $q(t)$ and $\phi(t)$. From this lagrangian we develop Hamiltonian, corresponding to two conjugate momentum $p_{q}$ and $p_{\phi}$. Since $\phi(t)$ is a cyclic co-ordinate thus $p_{\phi}$ is conserved. Solving Hamiltonian equation of motion for $q(t)$, we obtain cosmological relevant classical solutions. In this paper we studied the behaviour of certain cosmological parameter at late time epoch. Scale factor $a(t)$ evolves exponentially at late time cosmology. Energy density $\rho$ remains constant at late time cosmology, which can be identified as dark energy. At $t\rightarrow\infty$, equation of state parameter becomes $\omega\approx-1$. Thus satisfies necessary condition for dark energy. At late time cosmology, deceleration parameter $q_{0}\approx-1$, that satisfies an accelerated expansion of the universe.\\
Thus we get all the necessary relevant cosmological parameters from Hamilton's equation which predicts an accelerated expansion of the universe driven by negative pressure known as dark energy.


\begin{thebibliography}{}

\bibitem{Ref1} S.Perlmutter, et al, Nature. \textbf{391}, 51 (1998) .


\bibitem{Ref2} S.Perlmutter, et al, Astrophys.J. \textbf{517}, 565 (1999) .

\bibitem{Ref3} A.G.Riess, et al, Astrophys.J. \textbf{116}, 1009 (1998) .

\bibitem{Ref4} A.G.Riess, et al, Astrophys.J. \textbf{659}, 98 (2007) .

\bibitem{Ref5} R.R.Caldwell, R.Dave, P.J.Steinhardt, Phys.Rev.Lett. \textbf{80}, 1582 (1998) .

\bibitem{Ref6} W.M.Wood-Vesey, et al, Astrophys.J. \textbf{666}, 694 (2007) .


\bibitem{Ref7}  B.P.Schimdt, et al, Astrophys.J. \textbf{507}, 46 (1998) .

\bibitem{Ref8} N.W.Halverson, et al, Astrophys.J. \textbf{568}, 38 (2002) .

\bibitem{Ref9} E.Komatsu, et al, Astrophys.J.Suppl. \textbf{192}, 18 (2011) .

\bibitem{Ref10} C.L.Bennett, et al, Astrophys.J.Suppl. \textbf{148}, 1 (2003) .
 
\bibitem{Ref11} D.N.Spergel, et al, Astrophys.J.Suppl. \textbf{148}, 175 (2003) .


\bibitem{Ref12} P.A.R.Ade, et al, Astron.Astrophys. \textbf{A1},571 (2014) .

\bibitem{Ref13} N.Aghanim, et al, Astron.Astrophys. \textbf{Special issue} (2020) .

\bibitem{Ref14} M.Tegmark, et al, Phys.Rev.D. \textbf{69}, 103501 (2004) .

\bibitem{Ref15} C.A.Picon, T.Damour, V.Mukhanov, Phys.Lett.B. \textbf{458}, 209 (1999) .

\bibitem{Ref16} J.Garriga, V.Mukhanov, Phys.Lett.B. \textbf{458}, 219 (1999) .

\bibitem{Ref17} C.A.Picon, V.Mukhanov, P.J.Steinhardt, Phys.Rev.Lett. \textbf{85}, 4438 (2000) .

\bibitem{Ref18} T.Chiba, T.Okabe, M.Yamaguchi, Phys.Rev.D. \textbf{62}, 023511 (2000) .

\bibitem{Ref19} T.Chiba, Phys.Rev.D. \textbf{66}, 063514 (2002) .
 
\bibitem{Ref20} C.A.Picon, V.Mukhanov, P.J.Steinhardt, Phys.Rev.D. \textbf{63}, 103510 (2001) .

\bibitem{Ref21} L.P.Chimento, A.Feinstein, Mod.Phys.Lett.A. \textbf{19}, 761 (2004) .

\bibitem{Ref22} L.P.Chimento, Phys.Rev.D. \textbf{69}, 123517 (2004) .
 
\bibitem{Ref23} R.J.Scherrer, Phys.Rev.Lett.\textbf{ 93}, 011301 (2004) .

\bibitem{Ref24} D.Gangopadhyay, S.Mukherjee, Phys.Lett.B \textbf{665}, 121(2008) .  

\bibitem{Ref25} D.Gangopadhyay, S.Mukherjee, Grav.Cosmol.\textbf{ 17}, 349 (2011) .  

\bibitem{Ref26} L.P.Chimento, M.Forte, Phys.Rev.D \textbf{73}, 063502 (2006).

\bibitem{Ref27} K.Enqvist, Gen.Rel.Grav.\textbf{40}, 451 (2008) .

\bibitem{Ref28} C.H.Chuang,	Class.Quant.Grav. \textbf{ 25}, 175001 (2008) . 





 



\end{thebibliography}
\end{document}